\documentclass[prc,preprint,superscriptaddress,showpacs,amssymb,amsmath,amsfonts,aps]{revtex4}
\usepackage{graphicx}
\usepackage{epsfig}
\usepackage{dcolumn}

\begin{document}
\begin{flushright}
\preprint{JLAB-PHY-08-4}
\end{flushright}
\medskip

\title{Ratios of $^{15}$N/$^{12}$C and $^{4}$He/$^{12}$C 
Inclusive Electroproduction Cross Sections in the Nucleon Resonance Region}

\newcommand*{\ANL}{Argonne National Laboratory,  Argonne, Illinois 60439}
\affiliation{\ANL}
\newcommand*{\ASU}{Arizona State University, Tempe, Arizona 85287-1504}
\affiliation{\ASU}
\newcommand*{\UCLA}{University of California at Los Angeles, Los Angeles, California  90095-1547}
\affiliation{\UCLA}
\newcommand*{\CSU}{California State University, Dominguez Hills, Carson, CA 90747}
\affiliation{\CSU}
\newcommand*{\CMU}{Carnegie Mellon University, Pittsburgh, Pennsylvania 15213}
\affiliation{\CMU}
\newcommand*{\CUA}{Catholic University of America, Washington, D.C. 20064}
\affiliation{\CUA}
\newcommand*{\SACLAY}{CEA-Saclay, Service de Physique Nucl\'eaire, F91191 Gif-sur-Yvette, France}
\affiliation{\SACLAY}
\newcommand*{\CNU}{Christopher Newport University, Newport News, Virginia 23606}
\affiliation{\CNU}
\newcommand*{\UCONN}{University of Connecticut, Storrs, Connecticut 06269}
\affiliation{\UCONN}
\newcommand*{\ECOSSEE}{Edinburgh University, Edinburgh EH9 3JZ, United Kingdom}
\affiliation{\ECOSSEE}
\newcommand*{\FAI}{Fairfield University, Fairfield, CT 06824}
\affiliation{\FAI}
\newcommand*{\FIU}{Florida International University, Miami, Florida 33199}
\affiliation{\FIU}
\newcommand*{\FSU}{Florida State University, Tallahassee, Florida 32306}
\affiliation{\FSU}
\newcommand*{\GWU}{The George Washington University, Washington, DC 20052}
\affiliation{\GWU}
\newcommand*{\ECOSSEG}{University of Glasgow, Glasgow G12 8QQ, United Kingdom}
\affiliation{\ECOSSEG}
\newcommand*{\ISU}{Idaho State University, Pocatello, Idaho 83209}
\affiliation{\ISU}
\newcommand*{\INFNFR}{INFN, Laboratori Nazionali di Frascati, 00044 Frascati, Italy}
\affiliation{\INFNFR}
\newcommand*{\INFNGE}{INFN, Sezione di Genova, 16146 Genova, Italy}
\affiliation{\INFNGE}
\newcommand*{\ORSAY}{Institut de Physique Nucleaire ORSAY, Orsay, France}
\affiliation{\ORSAY}
\newcommand*{\ITEP}{Institute of Theoretical and Experimental Physics, Moscow, 117259, Russia}
\affiliation{\ITEP}
\newcommand*{\JMU}{James Madison University, Harrisonburg, Virginia 22807}
\affiliation{\JMU}
\newcommand*{\KYUNGPOOK}{Kyungpook National University, Daegu 702-701, South Korea}
\affiliation{\KYUNGPOOK}
\newcommand*{\MIT}{Massachusetts Institute of Technology, Cambridge, Massachusetts  02139-4307}
\affiliation{\MIT}
\newcommand*{\UMASS}{University of Massachusetts, Amherst, Massachusetts  01003}
\affiliation{\UMASS}
\newcommand*{\MOSCOW}{Moscow State University, General Nuclear Physics Institute, 119899 Moscow, Russia}
\affiliation{\MOSCOW}
\newcommand*{\UNH}{University of New Hampshire, Durham, New Hampshire 03824-3568}
\affiliation{\UNH}
\newcommand*{\NSU}{Norfolk State University, Norfolk, Virginia 23504}
\affiliation{\NSU}
\newcommand*{\OHIOU}{Ohio University, Athens, Ohio  45701}
\affiliation{\OHIOU}
\newcommand*{\ODU}{Old Dominion University, Norfolk, Virginia 23529}
\affiliation{\ODU}
\newcommand*{\PITT}{University of Pittsburgh, Pittsburgh, Pennsylvania 15260}
\affiliation{\PITT}
\newcommand*{\RPI}{Rensselaer Polytechnic Institute, Troy, New York 12180-3590}
\affiliation{\RPI}
\newcommand*{\RICE}{Rice University, Houston, Texas 77005-1892}
\affiliation{\RICE}
\newcommand*{\URICH}{University of Richmond, Richmond, Virginia 23173}
\affiliation{\URICH}
\newcommand*{\SCAROLINA}{University of South Carolina, Columbia, South Carolina 29208}
\affiliation{\SCAROLINA}
\newcommand*{\JLAB}{Thomas Jefferson National Accelerator Facility, Newport News, Virginia 23606}
\affiliation{\JLAB}
\newcommand*{\UNIONC}{Union College, Schenectady, NY 12308}
\affiliation{\UNIONC}
\newcommand*{\VT}{Virginia Polytechnic Institute and State University, Blacksburg, Virginia   24061-0435}
\affiliation{\VT}
\newcommand*{\VIRGINIA}{University of Virginia, Charlottesville, Virginia 22901}
\affiliation{\VIRGINIA}
\newcommand*{\WM}{College of William and Mary, Williamsburg, Virginia 23187-8795}
\affiliation{\WM}
\newcommand*{\YEREVAN}{Yerevan Physics Institute, 375036 Yerevan, Armenia}
\affiliation{\YEREVAN}
\newcommand*{\NOWJLAB}{Thomas Jefferson National Accelerator Facility, 
Newport News, Virginia 23606}
\newcommand*{\NOWOHIOU}{Ohio University, Athens, Ohio  45701}
\newcommand*{\NOWUNH}{University of New Hampshire, Durham, New Hampshire 03824-3568}
\newcommand*{\NOWUMASS}{University of Massachusetts, Amherst, Massachusetts  01003}
\newcommand*{\NOWMOSCOW}{Moscow State University, General Nuclear Physics Institute, 119899 Moscow, Russia}
\newcommand*{\NOWMIT}{Massachusetts Institute of Technology, Cambridge, Massachusetts  02139-4307}
\newcommand*{\NOWURICH}{University of Richmond, Richmond, Virginia 23173}
\newcommand*{\NOWODU}{Old Dominion University, Norfolk, Virginia 23529}
\newcommand*{\NOWCUA}{Catholic University of America, Washington, D.C. 20064}
\newcommand*{\NOWGEISSEN}{Physikalisches Institut der Universit\"{a}t Giessen, 35392 Giessen, Germany}
\newcommand*{\NOWLANL }{ Los Alamos National Laboratory, Los Alamos, New Mexico 87545}


\author{P.E.~Bosted}
     \email{bosted@jlab.org}
     \thanks{Corresponding author.}
\affiliation{\JLAB}
\author{R.~Fersch}
\affiliation{\WM}
\author {G.~Adams} 
\affiliation{\RPI}
\author {M.~Amarian}
\affiliation{\ODU}
\author {S.~Anefalos}
\affiliation{\INFNFR}
\author {M.~Anghinolfi} 
\affiliation{\INFNGE}
\author {G.~Asryan} 
\affiliation{\YEREVAN}
\author {H.~Avakian} 
\affiliation{\INFNFR}
\affiliation{\JLAB}
\author {H.~Bagdasaryan} 
\affiliation{\YEREVAN}
\affiliation{\ODU}
\author {N.~Baillie} 
\affiliation{\WM}
% ...
\collaboration{The CLAS Collaboration}
     \noaffiliation


%
%
\date{\today}



\begin{abstract}
The $(W,Q^2)$-dependence of the 
ratio of inclusive electron scattering  cross 
sections for
$^{15}$N/$^{12}$C was determined in the kinematic range 
$0.8<W<2$ GeV and $0.2<Q^2<1$ GeV$^2$ using 2.285 GeV electrons
and the CLAS detector at Jefferson Lab. The ratios exhibit
only slight resonance structure, in agreement with a 
simple phenomenological model and an extrapolation of DIS
ratios to low $Q^2$.
Ratios of $^4$He/$^{12}$C using 1.6 to 2.5 GeV electrons 
were measured with very high statistical precision, and were
used to  correct for He in the N and C targets. The $(W,Q^2)$
dependence of the $^4$He/$^{12}$C ratios is in good agreement with
the phenomenological model, and exhibit significant resonance
structure centered at  $W=0.94$, 1.23 and 1.5 GeV.
\end{abstract}
\maketitle

\section{Introduction}
While the internal structure of the proton has received enormous
attention over the past decades, the structure of its isospin
partner, the neutron, has not been studied as intensively.
The response of the neutron to electromagnetic probes is
not only of interest to characterize the structure of the nucleon,
but is also of considerable practical application, in particular
when the neutron is embedded in a nucleus. 

The application that
motivated the present study is the need to understand the
cross sections for electron scattering from $^{15}$N in
experiments using ammonia ($^{15}$NH$_3$ or $^{15}$ND$_3$) as
a source of polarized protons or deuterons. In experiments using
polarized ammonia to measure the spin structure functions
$g_1$ and $g_2$, data are also taken using a carbon  target,
so that in practice the ``dilution'' from unpolarized materials
(i.e. relative ratio of counts from $^{15}$N) can be determined
from good fits to the ratios $^{15}$N/$^{12}$C and measured ratios of
carbon to proton or deuteron cross sections. Polarized ammonia
targets are normally immersed in a bath of liquid He, so it
is also important to study $^4$He/$^{12}$C.

Another area where precision knowledge of the differences in
proton and neutron structure is of increasing practical importance
is in the field of neutrino scattering. The targets in these
low-rate experiments are normally made of heavy materials such
as iron to maximize count rates. The structure of the neutron
in iron has already been shown to be important to the interpretation
of the NuTeV experiment~\cite{nutev}. Precision knowledge 
of lepton-nucleon scattering from nuclei will be of particular
importance to the interpretation of planned neutrino
oscillation experiments~\cite{neutrino}.

A considerable body of data~\cite{E139} for inelastic lepton-nucleon
scattering from nuclei exists for a variety of nuclei in the 
deep-inelastic (DIS) region (missing mass $W>2$ GeV, 4-momentum transfer 
$Q^2>1$ GeV$^2$), where one expects the $n/p$ ratio in a nucleus
to be very similar to that determined from $d/p$ ratios. 
For $W<2$ GeV, 
the effects of prominent nucleon resonances, 
nucleon Fermi motion, Meson Exchange Currents (MEC)
and Final State Interactions (FSI) in larger nuclei
are of greater magnitude than in the case of deuterium,
so one can no longer simply use $d/p$ ratios to account for
the neutron excess in nuclei such as $^{15}$N, or Fe.  
Comparisons of nuclei with similar atomic number, but differing
ratios of neutrons to protons, are relatively scant in the
nucleon resonance region~\cite{nuc2}. The measurements of 
$^{15}$N/$^{12}$C over a wide range of $(W,Q^2)$ in the nucleon
resonance region from the present experiment were designed
to help address this situation. An interesting by-product
is a large body of high-statistical precision ratios of
$^4$He/$^{12}$C with a common systematic error.


\section{Experiment}
The present results were obtained as part of the 
Eg1b experiment~\cite{Eg1b} at Jefferson Lab, conducted in 2001.
The bulk of the experiment used polarized ammonia targets~\cite{Target}
to study electron scattering from 
polarized hydrogen and deuterium.  The ammonia content was dominated
by  $^{15}$N and was bathed in liquid He.  
Therefore we measured scattering from solid  $^{15}$N, 
carbon and empty targets, all bathed in liquid He, in order to understand 
and account for the unpolarized ``dilution factor''.
This article reports
on the results obtained with a 2.285 GeV electron beam
for the nitrogen to carbon ratios, and 
beam energies of 1.603, 1.721 and 2.285 GeV for the empty to carbon
target ratios. 

Electrons scattered at angles
between 10 and 45 degrees were detected in CLAS~\cite{CLAS}.
Experimental targets for $^{15}$N and $^{12}$C data consisted of
frozen $^{15}$N material inside a thin ($\sim$1.7 cm) target cup
and an amorphous carbon slab, respectively. Both were 
immersed in an approximately  1.7-cm-long, 
1 K LHe bath between two thin aluminum and Kapton
windows. The vertex tracking resolution was not sufficiently
precise to distinguish between events scattering from the windows,
the LHe bath, or the central nitrogen and carbon targets. 
A third target (MT) contained only LHe and the thin windows.
The mass thicknesses of all  target materials are listed 
in Table \ref{materials:table}. For modeling purposes, Kapton is
approximated as equivalent to $^{12}$C. While the thickness
of the carbon slab was relatively well known, the thickness of
LHe and nitrogen were relatively poorly determined, due to the
difficulties of working with targets at 1 K. 

\begin{table}[htbp]  \centering
\begin{tabular}{|c|c|c|c|}
\hline
Material & Targets & $t_m$ (g/cm$^2$) & $t_r$ (r.l.)\\
\hline
\hline
$^{12}$C & C & 0.501$\pm 0.005$ & 1.17\% \\
\hline
$^{15}$N & N & 0.56$\pm 0.03$ & 1.4\%\\
\hline
$^4$He & C, N, MT & 0.25$\pm 0.01$ & 0.2\% \\
\hline
Al & C, N, MT  & 0.045$\pm 0.002$ & 0.018\% \\
\hline
Kapton & C, N, MT & 0.072$\pm0.002$ & 0.017\% \\
\hline
\end{tabular}
\vspace{1cm}
\caption{Mass thickness $t_m$ and percentage of radiation length (r.l.) 
$t_r$ of materials  in each of the targets used in this experiment. 
}
\label{materials:table}
\end{table}

Scattered electrons were principally identified 
(and distinguished from pions)
by a Cherenkov threshold detector 
and electromagnetic shower calorimeter. 
Their momenta were determined by drift chambers 
and time-of-flight tracking 
in a (superconducting torus) magnetic field, 
with a trigger threshold of 0.3 GeV. Kinematically complete
elastic $ep$ and inelastic $ep\pi^+\pi^-$ events from separate NH$_3$
scattering data were used for calibration of the momentum
scale~\cite{Eg1banal}. Additional details on the beam, detectors, 
calibrations, and data analysis can be found in Ref.~\cite{Eg1banal}.

Ratios of count rates, normalized to total (live-time gated) incident
electron charge, were calculated for the nitrogen and carbon
target scattering events, and binned  in 
$Q^2$ and final state invariant mass $W$. All $^{15}$N/$^{12}$C
 ratios were for a 
beam energy of 2.285 GeV.
Similar ratios were also calculated between the carbon and
empty targets, for the determination of the $^{12}$C/$^4$He
cross-section ratios. In this case, the ratios at each of three
beam energies, 1.603, 1.721 and 2.285 GeV, were taken individually,
and the results then averaged. A $\chi^2$ test showed the three
sets of ratios to be statistically compatible. Small corrections
were made for pions mis-identified as electrons, and for electrons
from pair-symmetric decays of $\pi^0$ mesons (measured by 
reversing the CLAS torus polarity). Systematic errors in the cross
section ratios from these
corrections are generally very small, with a maximum of  0.5\% at the
highest $W$. 

Radiated cross-section models accounting for the mass and radiation
length thicknesses of
all materials within the targets were then fit
to the count ratio measurements. The two fit parameters were the
thickness of the nitrogen target and the length of the LHe.
Internal and external radiative effects were accounted for using
the formalism of Mo and Tsai~\cite{tsai}. The treatment of
radiative effects required modeling of nuclear elastic, quasi-elastic,
and inelastic cross sections from all nuclei in the targets.
The calculations used the nuclear charge
radii of Ref.~\cite{vries} for the elastic form factors 
(evaluated in the shell model), and the super-scaling
model of Ref.~\cite{Donnelly} for the quasi-elastic cross
sections with the nucleon elastic form factor parameterization
in Ref.~\cite{ne11fit} and the $y$-scaling function of 
Ref.~\cite{Amaro} for both the longitudinal and transverse
cross sections. The Donnelly model~\cite{Donnelly} has two
parameters (loosely related to Fermi broadening and 
average binding energy) that differ from nucleus to nucleus: we used
$(k_F,E_s)=(0.170,0.015)$ GeV for $^4$He and 
$(k_F,E_s)=(0.228,0.020)$ GeV for carbon and nitrogen. 
A Pauli-suppression correction at low $Q^2$ used 
the prescription of Mo and Tsai~\cite{tsai}.
Inelastic scattering was modeled using
a Fermi-convolution method of ``smearing'' 
free proton~\cite{F1pfit} and 
neutron~\cite{D2fit} cross sections. 
The smearing was done in a manner similar to that described
for the deuteron in Ref.~\cite{D2fit}. A detailed description,
with the numerous formulas and parameters involved, is 
beyond the scope of this article. However, FORTRAN computer code
for both the quasi-elastic and inelastic models used in this
paper is available~\cite{Bostedfit}.
Systematic errors were determined
by making reasonable variations in the cross section models and
target parameters.

The contributions from the LHe in the carbon and nitrogen targets
were subtracted using the radiated cross section global fit, 
as were the contributions from the aluminum and Kapton windows
in all three targets. Born-level ratios were then determined from
the background-subtracted ratios, using the calculated ratio of Born and
radiated cross sections for each target material. 

\section{Results for $^{4}$He/$^{12}$C}

\begin{figure}[hbt]
\centerline{\epsfig{file=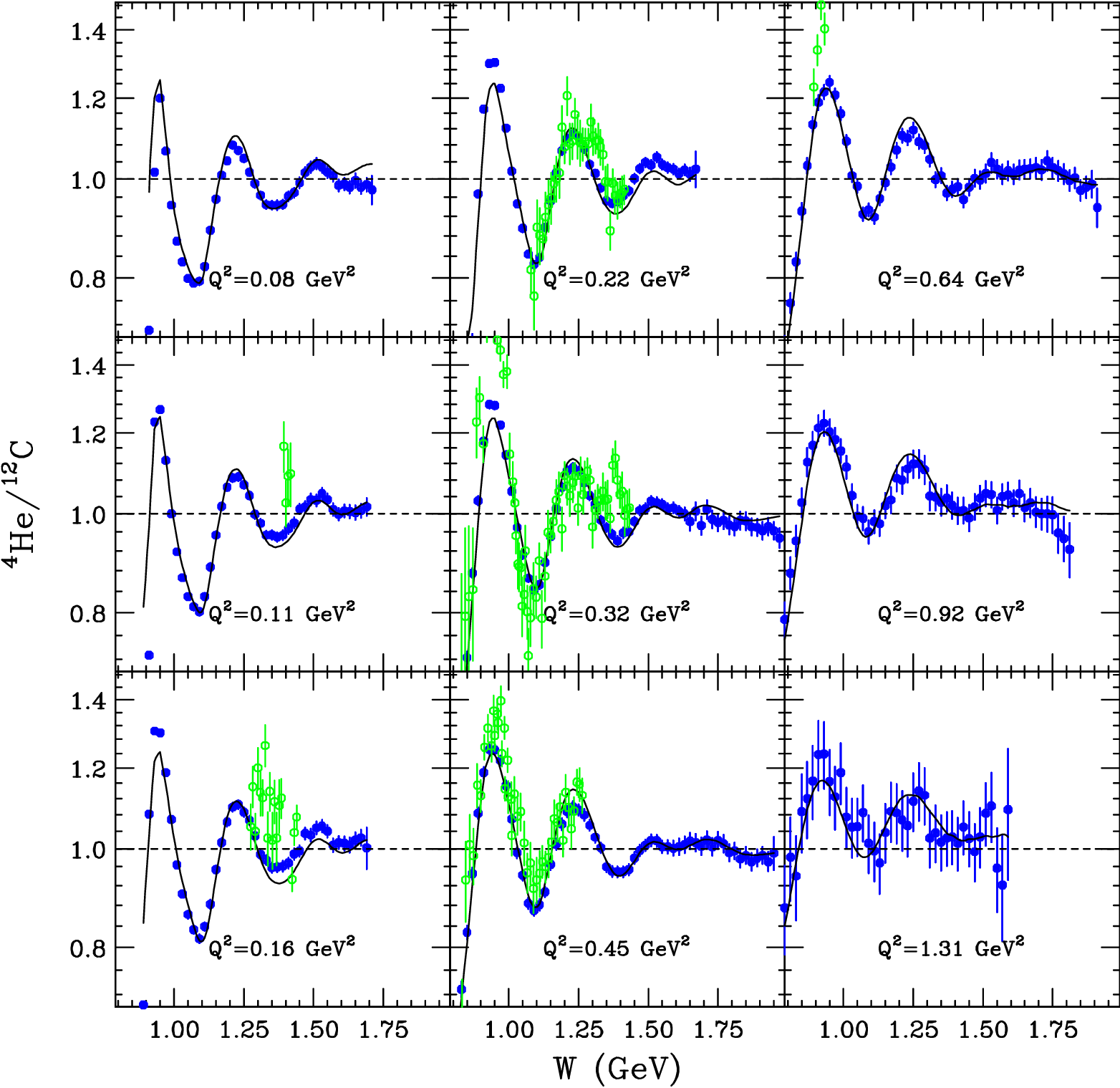,width=5.5in}}
\caption{(color online) Extracted ratios of  $^{4}$He to $^{12}$C 
cross sections (per nucleon) from this
experiment (solid blue circles), using beam energies of 1.6 to 2.2 GeV,
showing statistical and point-to-point systematic errors added in
quadrature.  The overall normalization error is  4\%. 
The solid curves show the ratio of model cross sections.
The green open circles are from SLAC experiment NE5~\protect{\cite{NE5}}.}
\label{fig:he}
\end{figure}

 The beam-energy averaged 
ratios of Born-level cross sections
for $^4$He/$^{12}$C are 
shown as the solid circles in Fig.~\ref{fig:he}.
Numerical results are available in the CLAS data base~\cite{CLASDB}.
It can be seen that the results are
generally in good agreement with $(W,Q^2)$ dependence of
the ratios from the Born-level inelastic cross section model that we
used for radiative corrections and background subtractions
(solid curves). This validates the use of this
model in correcting for the LHe contributions to the carbon 
and nitrogen targets. At moderate $Q^2$, the
present ratios are reasonably consistent with previous data
taken at a scattering angle of 37 degrees and beam energies of
0.9 to 1.2 GeV at SLAC~\cite{NE5} (open circles). The high
statistical precision of the present data is useful for 
constraining future fits to the kinematic dependence of helium
and carbon cross sections. In particular, the widths of the
peaks near $W=0.94$ GeV (from quasi-elastic scattering) and 
$W=1.23$ GeV (from excitation of the $\Delta(1232)$ resonance),
are very sensitive to the difference in average Fermi motion 
between helium and carbon (we used 180 MeV and 225 MeV,
respectively, for the Fermi smearing parameter $k_F$ in our model). 
The depth of the dip between these
two peaks is sensitive to possible differences in MEC and FSI
(we assumed no difference in our model). 
A slight peak in the ratios near $W=1.5$ GeV is also evident
in both the data and model at lower values of $Q^2$.

\section{Results for $^{15}$N/$^{12}$C}
\begin{figure}[hbt]
\centerline{\epsfig{file=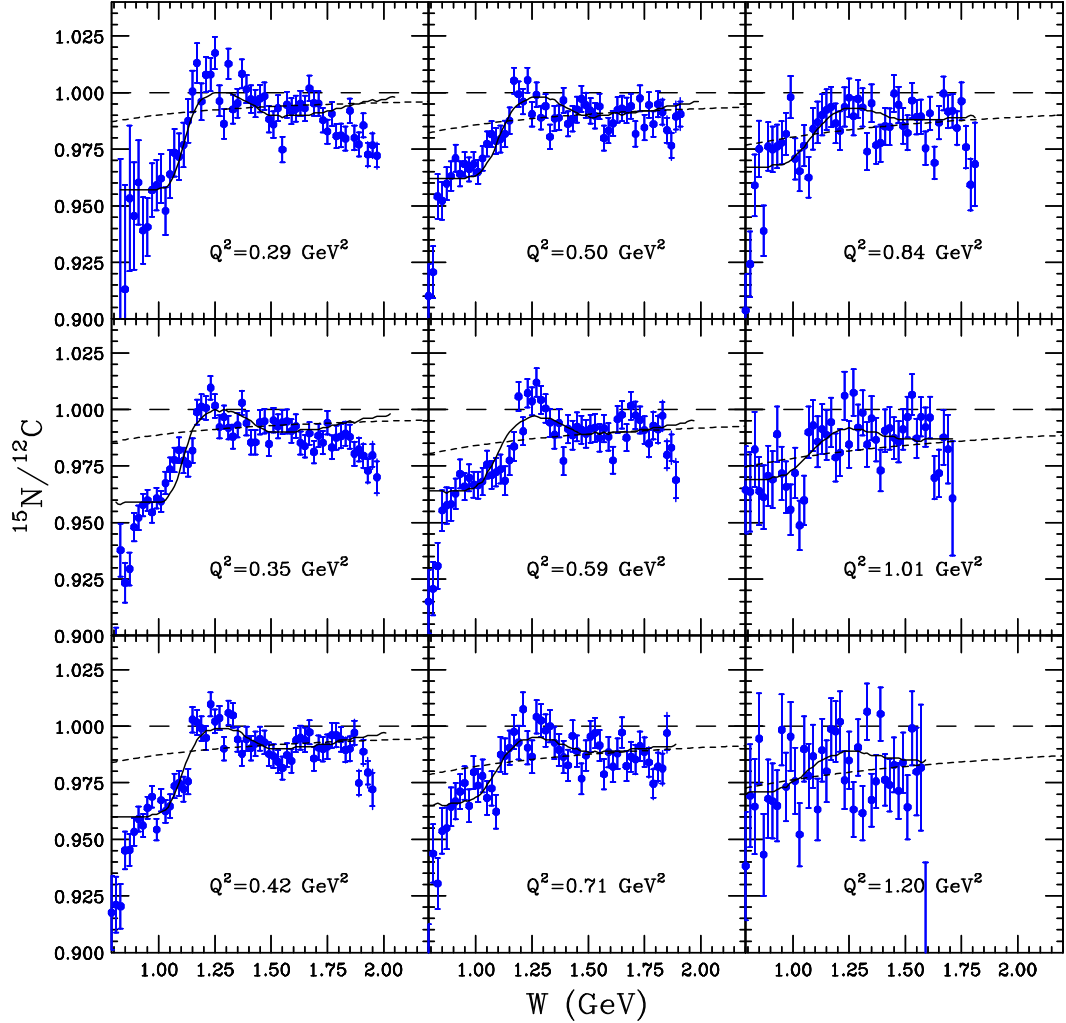,width=5.5in}}
\caption{(color online) Extracted ratios of cross sections per nucleon for 
pure  $^{15}$N to $^{12}$C from this
experiment (blue solid circles). The inner error bars show
statistical errors only, while the outer bars include 
point-to-point systematic errors added in quadrature. 
The overall normalization error
is 6\%. The solid curves show ratios of the model 
cross sections used for radiative corrections and
background subtraction. 
The dashed curves were generated using
$\sigma_n/\sigma_p=(1 - 0.8\xi)$. The long dashed lines are
plotted at unity, for reference. }
\label{fig:b}
\end{figure}

The ratios of extracted  Born cross sections per nucleon
for $^{15}$N/$^{12}$C are plotted in Fig.~\ref{fig:b} as a function
of invariant mass $W$ for nine bins in $Q^2$ and a beam energy
of 2.285~GeV. Point-to-point systematic errors (included in the
outer error bars) are relatively small. 
Due to the uncertainties in the target material thicknesses,
there is an overall normalization error of 6\%.
Numerical results are available in the CLAS data base~\cite{CLASDB}.

In general, there is good agreement with the cross section model ratios
used for radiative and background corrections, shown as the solid curves.
The pronounced dips in the ratios near the quasi-elastic region
(W=0.94 GeV)  ``level off'' with increasing $Q^2$, 
due to the increasing contributions from inelastic scattering, and 
the increasing ratio of neutron to proton elastic form factors.
The ratios in the resonance region ($W>1.1$ GeV)
show only slight resonant structure. At lower $Q^2$,
both data and the model show enhanced ratios near the $\Delta(1232)$
peak, where the neutron to proton ratio is expected to approach
unity due to the isovector nature of this resonance, while
the non-resonant background has a smaller ratio~\cite{D2fit}.
The enhancement near $W=1.23$ GeV in the model is not due to a difference
in Fermi motion, because the same average Fermi momentum was
assumed in the model for carbon and nitrogen. 
A slight dip near the $S_{11}(1535)$ resonance is possibly
evident.  At higher $Q^2$, Fermi-smearing effects become
more significant, and all indications of resonant structure disappear. 

On average, the ratios tend to decrease with increasing
$Q^2$ at fixed $W$, corresponding to larger values of the
Bjorken scaling variable $x$. In deep-inelastic scattering, 
$\sigma_n/\sigma_p$ is approximately given by $(1-0.8x)$ \cite{SMC}. 
To approximately take into account target-mass effects, 
$x$ was replaced with the Nachtmann~\cite{Nachtmann1975} 
scaling variable $\xi \equiv 2x/(1+\sqrt{1+4M^2x^2/Q^2})$.
The dashed curve generated defined by 
$\sigma(^{15}$N$)/\sigma(^{12}$C)$ = 1 - (1-\sigma_n/\sigma_p)
/15(1+\sigma_n/\sigma_p)$ using
$\sigma_n/\sigma_p=(1-0.8\xi)$, shown  
in Fig.~\ref{fig:b}, is a remarkably good approximation of
the data, especially at the low 
$Q^2$ values of this experiment, where additional higher  twist
effects might be expected to play a
role. This is particularly true if one averages over
the quasi-elastic and $\Delta$ resonance regions~\cite{Isgur}; in
which case the resultant curve matches the Nachtmann-scaled
extrapolation into the DIS region.
This appears to perhaps be yet another manifestation of
quark-hadron duality, the fulfillment of which implies
the marked absence or cancellation of 
higher-twist effects \cite{Ent}. That these higher twist, 
multi-parton contributions appear to cancel nearly completely in the
ratio is a noteworthy, if not unexpected phenomenon at these values of
$Q^2$.

\section{Summary}
We find that the $(W,Q^2)$-dependence of ratios of 
electroproduction cross sections for 
$^{15}$N/$^{12}$C and $^{4}$He/$^{12}$C can be remarkably
well-described by a simple model based on super-scaling in the 
quasi-elastic region, and simple Fermi-smearing 
in the nucleon resonance region,
even at $Q^2$ values as low as 0.1 GeV$^2$. Large oscillations
in the ratios of $^{4}$He/$^{12}$C peaked near $W=0.94$, 1.23,
and 1.5 GeV can be attributed to a difference in average Fermi
momentum. In contrast, little 
structure is seen in the ratios of $^{15}$N/$^{12}$C, 
except for small effects in the $\Delta(1232)$ region and
a decrease in the quasi-elastic region expected from the ratio
of neutron to proton form factors.  The new data can be used
to refine more detailed microscopic models of lepton-nucleon
scattering in the nuclear medium. 

Suitably averaged over $W$ (as for example as may occur naturally
with the use of a wide-band neutrino beam), the ratios of 
$^{15}$N/$^{12}$C and $^{4}$He/$^{12}$C bear a strong resemblance
to extrapolations of DIS models into the nucleon resonance
region. This observation might be used to simplify predictions
for neutrino oscillation experiments. It is also another
indication of the applicability of the concepts of quark-hadron
duality down to remarkably low values of $Q^2$.

\section{Acknowledgments}
We would like to acknowledge the outstanding efforts of the 
Accelerator, Target Group, and Physics Division staff that made
this experiment possible. 
This work was supported by the U.S. Department of Energy, the
Italian Istituto Nazionale di Fisica Nucleare, the U.S. National
Science Foundation, the French Commissariat \`a l'Energie Atomique 
and the Korean Engineering and Science 
Foundation. The Southeastern
Universities Research Association (SURA) operated the Thomas
Jefferson National Accelerator Facility for the United States
Department of Energy under contract DE-AC05-84ER40150.

\end{document}